\documentclass[10pt,twocolumn]{article}
\usepackage{epsfig,amsfonts,amsmath,graphicx,dcolumn,bm,amssymb,authblk,cite,epstopdf,times,fullpage,placeins}
\usepackage[hmargin=2.5cm,vmargin=3cm]{geometry}

\makeatletter
\def\@cite#1#2{$^{\mbox{\scriptsize #1\if@tempswa , #2\fi}}$}
\renewcommand{\section}{\@startsection {section}{1}{0pt}%
    {-6pt}{1pt}%
    {\bfseries}%
    }
\renewcommand\@biblabel[1]{#1.}
\makeatother
\renewenvironment{abstract}{%
    \setlength{\parindent}{0in}%
    \setlength{\parskip}{0in}%
    \bfseries%
    }{\par\vspace{-6pt}}
\newcommand{\spacing}[1]{\renewcommand{\baselinestretch}{#1}\large\normalsize}
\spacing{1}

\title{\textbf{Cavity ring-up spectroscopy for ultrafast sensing with optical microresonators}}
\author{Serge Rosenblum$^{1\dag}$, Yulia Lovsky$^{1\dag}$, Lior Arazi$^1$, Frank Vollmer$^2$ and Barak Dayan$^{1\star}$}

\begin{document}

\begingroup
\let\center\flushleft
\let\endcenter\endflushleft
\date{}
\maketitle
\endgroup
\let\thefootnote\relax\footnote{$^1$Department of Chemical Physics, Weizmann Institute of Science, 76100 Rehovot, Israel, $^2$Laboratory of Nanophotonics \& Biosensing,
 Max Planck Institute for the Science of Light, D-91058 Erlangen, Germany; $^\dag$These authors contributed equally to this work. $^\star$e-mail: barak.dayan@weizmann.ac.il}
\begin{abstract}
Spectroscopy of whispering-gallery mode (WGM) microresonators has become a powerful scientific tool, enabling detection of single viruses, nanoparticles, and even single molecules.
Yet the demonstrated timescale of these schemes has been limited so far to milliseconds or more.
Here we introduce a novel scheme that is orders of magnitude faster, capable of capturing complete spectral snapshots of WGM resonances at nanosecond timescales: cavity ring-up spectroscopy (CRUS).
Based on sharply-rising detuned probe pulses, CRUS combines the sensitivity of heterodyne measurements with the highest possible, transform-limited acquisition rate.
As a demonstration we capture spectra of microtoroid resonators at time intervals as short as $16$ ns, directly monitoring sub-microsecond dynamics of their optomechanical vibrations, thermorefractive response and Kerr nonlinearity.
CRUS holds promise for the study of fast biological processes such as enzyme kinetics, protein folding and light harvesting, with applications in other fields such as cavity QED and pulsed optomechanics.
\end{abstract}

\section*{Introduction}
\vspace{-0.5cm}
The remarkable sensitivity of WGM microresonators\cite{braginsky1989quality,vahala2003optical} stems from their small mode volume ($V\sim100\:\mu \textrm{m}^3$), which enhances the electric field associated with every photon, and their ultra-high quality factor ($Q\sim10^8$), which enables very long cavity lifetimes.
This leads to enhancement of the interaction with any object within the mode\cite{Vollmer@Keng_PNAS_2008,Zhu@Yang_NatPhotonics2009,lu2011,He@Yang_NatNano2011,arnold2013label} by a factor proportional to $Q/V$. As a result, the presence of even nanoscale objects within the evanescent field of the WGM can lead to detectable variations in its spectrum\cite{mazzei2007}. Specifically, these variations include broadening (due to shortening of the cavity lifetime)\cite{shao2013detection,truong2013}, splitting (due to induced cross-talk between its two counter-propagating modes)\cite{mazzei2007,Zhu@Yang_NatPhotonics2009} or a shift (due to modification of the optical path length)\cite{arnold2003shift,lu2011}.
The typical scale of all these spectral variations is of the order of a few MHz or more, in principle allowing their detection on a timescale of less than a microsecond.\\ \vspace{-1cm}

%
However, although there are fast methods to separately measure splitting\cite{He@Yang_NatNano2011,knittel2013back}, broadening\cite{truong2013} or a shift\cite{keng2007}, the only method for obtaining the full spectrum had so far been laser scanning.
Laser scanning is slow not only because of technical reasons. It is inherently slow since in order to accurately capture the cavity spectrum, the laser must scan over the cavity resonance (of width $2\kappa$) during a time long enough for the intracavity field to continuously maintain a steady state. The timescale for this duration is determined by the cavity lifetime $\tau$ (=$1/4\pi\kappa$), meaning that the scan rate needs to be significantly slower than $\kappa/\tau$. Accordingly, to acquire a spectral scan over a total bandwidth $B$, the total scan time needs to exceed $N\tau$, with $N=B/\kappa$ representing the effective number of frequency points acquired during the scan. Yet this scan time is not optimal:
the fundamental limit for the duration of such a spectral measurement is defined by the time-bandwidth uncertainty relation (or the Fourier limit), and is set by the desired spectral resolution ($\kappa$) to be roughly $\tau$.
In other words, a scan measurement is at least $N$ times slower than a Fourier-limited one.\\ \vspace{-1cm}

\begin{figure}[t]
\center
\includegraphics[width=0.5\textwidth]{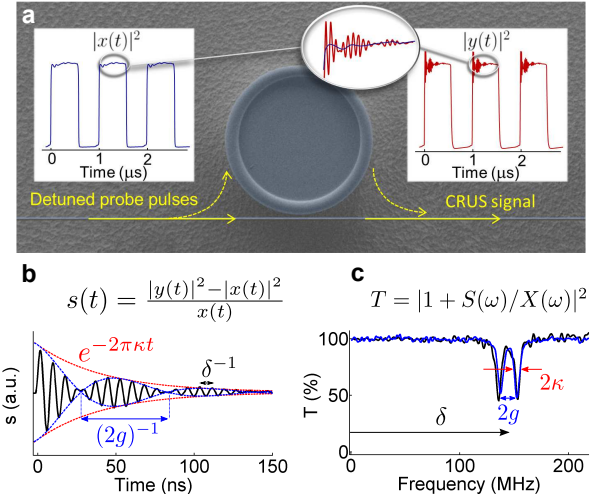}
\caption{\textbf{Outline of the principles of CRUS. a}, SEM-image of a tapered-fibre coupled microtoroid. Sharply-rising detuned probe pulses (left inset) lead to the buildup of a cavity field. As this weak field leaks back into the fibre, it interferes with the probe, resulting in a beating signal at the output (right inset). \textbf{b}, The ring-up signal $s(t)$. The detuning $\delta$ from the probe is exhibited by the fast oscillations, the resonance width $2\kappa$ is exhibited by the exponential decay envelope, and the slow beat note indicates the splitting of the resonance $2g$. \textbf{c}, Fourier transforming $s(t)$ allows retrieval of the complete spectrum (black curve), showing excellent agreement with the four orders of magnitude slower scan measurement (blue curve). Both spectra were taken with similar peak intracavity energy ($\sim 100$ fJ, set below the visible onset of nonlinear effects).\label{fig:crus}}
\end{figure}

Here we present CRUS as an ultrafast method for acquiring complete spectral snapshots of WGM resonators at the minimal timescale set by the Fourier limit. CRUS relies on the abrupt turn-on of far-detuned probe laser pulses ($\delta \gg 2\kappa$), that are coupled to the WGM microresonator (a microtoroid\cite{armani2003} in our case) through a tapered optical fibre\cite{Knight@Birks_OL_1997} (Fig.~\ref{fig:crus}a). At such a large detuning, most of the light is transmitted through the fiber without entering the cavity. However, the short rise time $t_r$ of the pulse corresponds to a broad spectral distribution of bandwidth $B\sim1/t_r$. Accordingly, if the cavity resonance lies within this bandwidth ($\delta<B$), this turn-on leads to the buildup of a small, transient field in the cavity. This field then radiates back into the fibre, where it interferes with the transmitted field, creating a “ring-up” signal. The spectrum can now be retrieved with a resolution of $\sim \kappa$ (or better) by acquiring this ring-up signal during a time of $\tau$ (or longer).\\ \vspace{-1cm}

CRUS therefore enables monitoring kinetics and events in the nanosecond regime, drastically expanding the capabilities demonstrated in previous schemes\cite{Vollmer@Keng_PNAS_2008,Zhu@Yang_NatPhotonics2009,lu2011,He@Yang_NatNano2011}, which have focused on slow detection of long-term events, such as binding of viruses or nanoparticles to the surface of the microresonator.\\ \vspace{-1cm}

This situation is reminiscent of the well-known method of cavity ring-down spectroscopy (CRDS), in which the field emitted from a cavity is directly measured by abruptly turning off a resonant probe\cite{okeefe1988}.
CRDS is widely used as a sensitive method for absorption spectroscopy of gaseous samples in Fabry-P\'{e}rot resonators. Since CRDS captures only the intensity of the ring-down signal, it provides no knowledge of the center frequency of the resonance, and therefore cannot be used to detect spectral shifts (nor does it enable reconstruction of asymmetric spectra). Accordingly, its application to WGM microresonators had so far been limited to Q-factor measurements\cite{armani2003}.\\ \vspace{-1cm}

In contrast, in CRUS the far-detuned probe, which continues in the fibre, serves as a local oscillator as it interferes with the emitted cavity field. The result is therefore a robust heterodyne measurement of both the amplitude and the phase of the cavity field, which allows unambiguous extraction of the complete spectrum, including its center frequency. The difference between CRUS and CRDS is therefore analogous to the difference between a hologram, which captures both the amplitude and the directionality of the object wavefront, to a picture, which captures only its intensity. In this view, the large detuning between the probe and the resonance is analogous to the large angular separation between the reference beam and the object in an off-axis hologram, which prevents overlap between positive and negative frequencies (the object and its conjugate) during the reconstruction.\\

\section*{Results}
\begin{figure*}[t!]
 \begin{minipage}{\textwidth}
             \includegraphics[width=\textwidth]{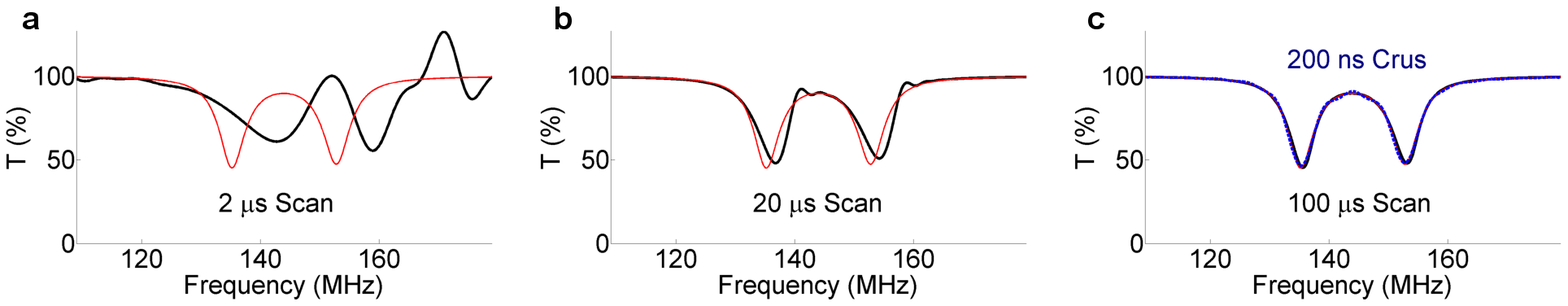}\quad
             \caption{\textbf{Comparison of scanning spectroscopy and CRUS.}  Simulated scan measurements (black curves) of the spectrum of a resonator identical to that of Fig. 1 (red curves),  with scan times of \textbf{a,} $2$ $\mu$s, \textbf{b,} $20$ $\mu$s and \textbf{c,} $100$ $\mu$s, assuming a total scan range of $500$ MHz (not entirely shown in figure). The blue curve in \textbf{c} shows the simulated result of a CRUS measurement using a $200$ ns ring-up signal. \vspace{.75cm} \label{fig:scan_vs_crus}}
 \end{minipage}

 \begin{minipage}{\textwidth}
\includegraphics[width=\textwidth]{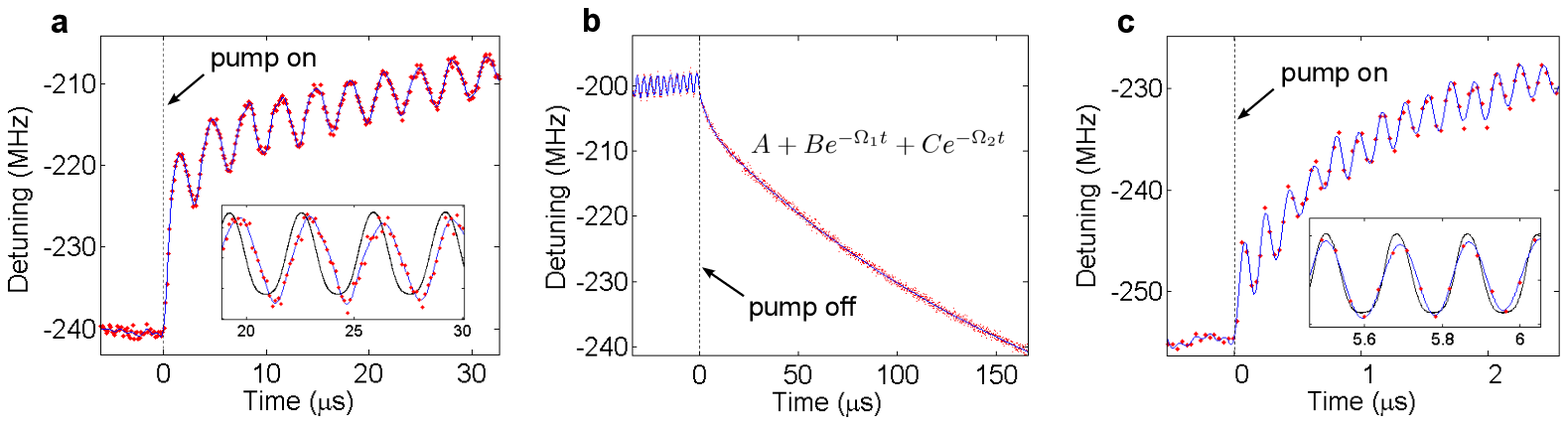}\quad
             \caption{\textbf{Thermal and Kerr nonlinearities measured by CRUS.} The graphs show the center frequency of the probed WGM (red dots, interpolation in blue) as extracted from CRUS spectra for: \textbf{a,} Thermal oscillations induced by the turn-on of a pump beam modulated at $300$ kHz. The oscillations are superimposed on a drifting central frequency which reflects the gradual increase in the microtoroid's average temperature.
            The spectra were captured every $200$ ns. The inset shows the phase lag between the induced oscillations and the pump (black curve). \textbf{b,} The relaxation of the resonant frequency following the turn-off of the pump. The cooling of the cavity can be separated into a slow decay (corresponding to a cutoff frequency of $\Omega_1 = 5.1$ kHz) and a fast decay (cutoff frequency $\Omega_2 = 210$ kHz). \textbf{c,} Kerr oscillations induced by the turn-on of a pump beam modulated at $5.5$ MHz. The spectra were captured every $37$ ns. The instantaneous response of the Kerr effect is exhibited by nearly zero phase shift with respect to the pump (inset).\label{fig:modulation}}
 \end{minipage}
\end{figure*}

\vspace{-0.5cm}
Figure \ref{fig:crus}b presents the measured ring-up signal, defined as $s(t)=\frac{\left|y\right|^2-\left|x\right|^2}{x}$, where $x$ and $y$ are the input and output fields, respectively. The main features of the WGM resonance already appear in this signal. Accordingly, $s(t)$ enables complete retrieval of the transmission spectrum $T(\omega)=|Y(\omega)/X(\omega)|^2$, which by the final value theorem is identical to that obtained by frequency scanning (where $X$,$Y$ are the Fourier transforms of $x$, $y$). This is done by noting that the output field results from interference between the input field and the emission from the cavity, i.e. $Y(\omega)= X(\omega)+X(\omega)H(\omega)$, where $H(\omega)$ is the cavity's transfer function. By Fourier transforming the ring-up signal $s(t)$ and using the fact that due to the large detuning $H(\omega)$ contains only positive frequencies, we get $H(\omega)=S(\omega)/X(\omega)$, finally yielding (see Methods)
\begin{eqnarray}
T(\omega)&=&|1+S(\omega)/X(\omega)|^2.\nonumber
\end{eqnarray}
Figure~\ref{fig:crus}c presents the spectrum obtained from a single-shot, $500$ ns pulse performed on a resonance with $Q=1\times10^8$ at $780$ nm, showing excellent agreement with the much slower ($5$ ms) single-shot scan measurement. The fundamental origin of the drastic difference between the speeds of CRUS and a regular scan is clearly illustrated in Fig.~\ref{fig:scan_vs_crus}, which simulates scans of the spectrum of Fig.~\ref{fig:crus}c at various speeds. The fact that CRUS is a transform-limited measurement whereas a scan measurement is not, is clearly exhibited by the ringing effects that appear at short scan times\cite{matone2000finesse}.\\ \vspace{-1cm}

The ability to rapidly capture the entire spectrum also makes CRUS more resilient to various noise mechanisms (such as amplitude noise) that typically occur at lower frequencies and can distort a scan measurement, which samples different parts of the spectrum at different times. Indeed, as evident from Fig.~\ref{fig:crus}c, the signal-to-noise ratio of CRUS is comparable to that of the much slower scan.
CRUS also has the inherent advantage of heterodyne measurements, which readily enable shot-noise limited sensitivities (as was demonstrated for heterodyne detection of cavity ring-down by Levenson \textit{et al.}\cite{levenson1998optical,levenson2000frequency}, and for probing of WGM nonlinear dynamics\cite{lin2008}). Sensitive detection of weak fields is extremely important with WGM microresonators, in which the intracavity power must be kept very low, to prevent the onset of undesired nonlinearities. \\ \vspace{-1cm}

%
As a first demonstration of the potential of CRUS for capturing ultrafast phenomena, we applied it to fully characterize the nonlinearities of the microtoroid.
For this purpose we used a modulated pump laser (see Methods) to induce spectral variations in the probed WGM resonance through thermal\cite{ilchenko1992}, Kerr\cite{rokhsari2005observation,vukovic2013} and optomechanical effects\cite{rokhsari2005radiation}.\\ \vspace{-1cm}

Figure~\ref{fig:modulation}a shows the response of the microtoroid to an abrupt turn-on of a $300$ kHz modulated pump, as exhibited by the variations in the center frequency of a $54$ MHz WGM resonance (full width at half maximum), captured by CRUS at intervals of $200$ ns. As demonstrated in previous works\cite{ilchenko1992,schliesser2006radiation}, for modulation frequencies below $\sim1$ MHz the nonlinearity of the microtoroid is dominated by its thermal response, which is governed by two processes with different timescales\cite{ilchenko1992,schliesser2006radiation}. The short timescale is set by the conduction rate of heat from the optical mode to the rest of the cavity, and the longer corresponds to dissipation of heat from the entire cavity to the silicon substrate through the pillar. The high rate of CRUS allows accurate extraction of these timescales directly from the cavity's temporal response to abrupt turn-off of the pump, as shown in Fig.~\ref{fig:modulation}b.\\ \vspace{-1cm}

The CRUS measurements also reveal a phase lag of  $50^\circ\pm1^\circ$ between the cavity oscillations and the pump modulation (inset of Fig.~\ref{fig:modulation}a), which is a result of the interplay between the two capacitive thermal mechanisms and the much weaker, frequency-independent Kerr nonlinearity.
By measuring this phase shift as a function of modulation frequency we find that the heating rates are $\approx6,700$ K/W/100 ns for the slow mechanism and $\approx700$ K/W/100 ns for the fast mechanism, in good agreement with theory\cite{schliesser2009cavity}.\\ \vspace{-1cm}

Figure~\ref{fig:modulation}c shows the cavity's center frequency response to abrupt turn-on of a pump laser modulated at $5.5$ MHz, extracted from spectral snapshots taken every $37$ ns. At this high frequency the nonlinearity of the cavity is dominated by the Kerr effect\cite{rokhsari2005observation,schliesser2006radiation} (superimposed on the slower thermal drift), with the corresponding instantaneous response exhibited by a nearly zero phase shift ($4^\circ\pm1^\circ$) with respect to the pump modulation (see inset). \\ \vspace{-1cm}

\begin{figure}[b!]
                \centering
                \includegraphics[width=0.4\textwidth]{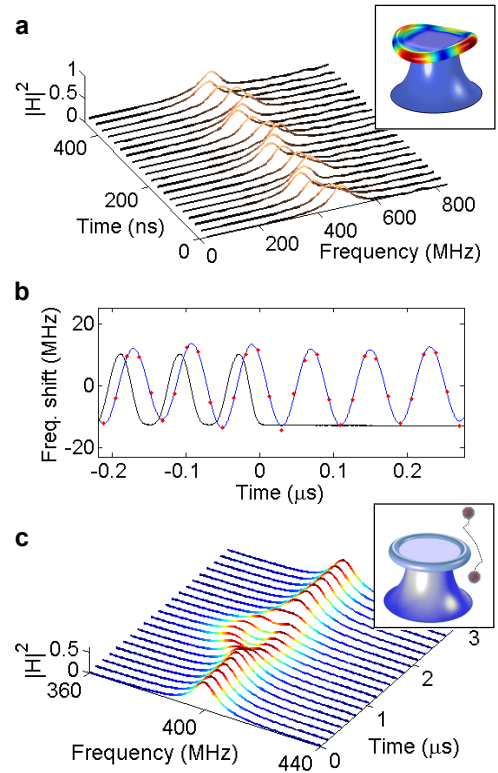}\quad
             \caption{\textbf{CRUS measurements of optomechanical oscillations. a,} Measured spectra of a microresonator oscillating mechanically at $8$ MHz, as captured by CRUS at intervals of $25$ ns. \textbf{b}, The center frequency shift (red dots, interpolation in blue) of a microresonator oscillating mechanically at $12.5$ MHz, extracted from CRUS taken every $16$ ns. The oscillations display a $\pi/2$ phase shift relative to the pump (black curve) and continue after the pump is turned off, as expected from a resonantly driven oscillator. \textbf{c,} Simulated CRUS performed every 150 ns, for a 75 nm radius nanoparticle with refractive index $n=1.5$ crossing the evanescent field of a 10 MHz WGM at a speed of 1 m/s, with distance of closest approach of 150 nm (measured from the particle center to the resonator surface).  \label{fig:oscillations}}
\end{figure}

As a final demonstration of the capabilities of CRUS we directly monitor optomechanical oscillations\cite{Kippenberg@Vahala_Science_2008,aspelmeyer2013,rokhsari2005radiation} of the microtoroid at a rate much faster than the oscillation frequency. Such pulsed measurements are of particular interest in the context of quantum non-demolition measurements of mechanical motion\cite{Vanner@Aspelmeyer_PNAS_2011}.
For this purpose we tuned the pump to excite the second order crown mode of the microtoroid (inset of Fig.~\ref{fig:oscillations}a).\\ \vspace{-1cm}

Figure~\ref{fig:oscillations}a presents the measured time-varying spectrum of a $90$ MHz WGM resonance, captured at intervals of $25$ ns. The graph clearly displays the variations of the resonance resulting from the oscillations of the microtoroid's circumference at $8$ MHz.
Figure~\ref{fig:oscillations}b displays the center frequency shift extracted from CRUS spectra of another microtoroid, in which the frequency of the second-order crown mode was $12.5$ MHz. Here the spectra were taken at intervals of $16$ ns. The extracted phase difference between the mechanical oscillations and the pump is $89^\circ\pm1^\circ$, in agreement with the expected value of $\pi/2$ for a resonantly driven oscillator. The resonant behavior of these oscillations is also exhibited by the fact that they continue after the pump is abruptly turned off. By measuring the decay rate of these oscillations we obtain a mechanical quality factor of $300$.
These results provide the first direct observation of optomechanical oscillations by continuous monitoring of the entire WGM spectrum - a task far beyond the ability of laser scanning.\\

\section*{Discussion}
\vspace{-0.5cm}
By acquiring maximal knowledge on the spectrum at the shortest possible timescale, CRUS introduces the promising field of optical sensing with WGM microresonators into the nanosecond regime. As a specific example, Fig.~\ref{fig:oscillations}c presents a numerical simulation of a nanoparticle with radius of $75$ nm flying by the microtoroid. As evident in this example, CRUS enables simultaneous observation of the shift, splitting and broadening induced by the interaction with the nanoparticle.
Looking forward, these abilities open the path for the study of ultrafast phenomena in various fields, such as micromechanical vibrations\cite{Vanner@Aspelmeyer_PNAS_2011,krause2012high}, cavity QED\cite{aoki2006} and fast biological processes such as enzyme kinetics\cite{yasuda2001resolution}, light harvesting\cite{perman1998energy} and protein folding\cite{mayor2003complete}.
\vspace{-0.5cm}

\section*{Methods}
\vspace{-0.5cm}
The probe laser (at 780 nm) was intensity modulated with a LiNbO$_3$ electro-optic modulator, yielding pulses with rise times typically below $1$ ns. The pulse length was set by the required spectral resolution, and the cavity lifetime was tuned to be significantly shorter by increasing the coupling to the tapered fibre.
To allow separation from the probe, the pump laser was tuned to a WGM resonance at 770 nm (different from the probed resonance) and set to be counterpropagating to the probe.\\ \vspace{-1cm}

Due to the large detuning, the term quadratic in the cavity emission is neglected in the retrieval algorithm, as it is much weaker than the probe. Additionally, it was assumed that the input field $x$ has a constant phase, and accordingly it was treated as real. This assumption typically holds except, in some cases, during the pulse rise. If the pulse rise time is significantly shorter than $\delta^{-1}$, this effect can still be neglected. Otherwise, accurate retrieval of the spectrum requires knowledge of this transient phase, which depends on the specific modulation technique used (in our case a Z-cut modulator with a chirp factor of $0.7$).

\vspace{-0.5cm}
\section*{Author Contributions}
\vspace{-0.5cm}
Y.L., S.R., and B.D. conceived the idea. Y.L. and S.R. conducted the experiments and analysed the data. B.D., Y.L., S.R. and F.V. wrote the manuscript. All authors contributed to the design and construction of the experiment, discussed the results and commented on the manuscript.
\vspace{-0.5cm}
\section*{Additional information}
\vspace{-0.5cm}
The authors declare no competing financial interests. Correspondence and requests for materials should be addressed to B.D.\end{document}